\documentclass[preprint,12pt,authoryear]{elsarticle}
\usepackage{natbib}

\usepackage{amssymb}
\usepackage{amsmath}
\usepackage{booktabs} 
\usepackage{caption}  
\usepackage{enumitem}
\usepackage{xcolor}
\usepackage{float}
\usepackage{tabularx}

\journal{}
\begin{document}
\begin{frontmatter}


\title{TUMLS: Trustful Fully Unsupervised Multi-Level Segmentation for Whole Slide Images of Histology}


\author{Walid Rehamnia, Alexandra Getmanskaya, Evgeniy Vasilyev, Vadim Turlapov}

\affiliation{organization={Lobachevsky State University of Nizhny Novgorod},
            addressline={Prospekt Gagarina, 23}, 
            city={Nizhny Novgorod},
            postcode={603022}, 
            state={Nizhny Novgorod},
            country={Russia}}

\begin{abstract}

Digital pathology, augmented by artificial intelligence (AI), holds significant promise for improving the workflow of pathologists. However, challenges such as the labor-intensive annotation of whole slide images (WSIs), high computational demands, and trust concerns arising from the absence of uncertainty estimation in predictions hinder the practical application of current AI methodologies in histopathology. To address these issues, we present a novel trustful fully unsupervised multi-level segmentation methodology (TUMLS) for WSIs.
TUMLS adopts an autoencoder (AE) as a feature extractor to identify the different tissue types within low-resolution training data. It selects representative patches from each identified group based on an uncertainty measure and then does unsupervised nuclei segmentation in their respective higher-resolution space without using any ML algorithms.
Crucially, this solution integrates seamlessly into clinicians’ workflows, transforming the examination of a whole WSI into a review of concise, interpretable cross-level insights. This integration significantly enhances and accelerates the workflow while ensuring transparency. We evaluated our approach using the UPENN-GBM dataset, where the AE achieved a mean squared error (MSE) of 0.0016. Additionally, nucleus segmentation is assessed on the MoNuSeg dataset, outperforming all unsupervised approaches with an F1 score of 77.46\% and a Jaccard score of 63.35\%. These results demonstrate the efficacy of TUMLS in advancing the field of digital pathology.

\end{abstract}




\end{frontmatter}

\section{Introduction}
Whole slide imaging has become an integral part of modern digital pathology, enabling detailed, high-resolution analysis of tissue samples \citep{wright2013digital} \citep{gu2018multi}. However,
the computational cost associated with processing gigapixel WSI data can be substantial and often pose significant challenges for effective segmentation, a critical step in numerous clinical and research applications \citep{ronneberger2015u, gu2018multi}. Additionally, most of the outperforming methods in medical image segmentation rely on extensive pixel-level annotations, which are laborious to obtain and often fail to generalize well to diverse clinical samples \citep{peng2021medical}, which goes against the primary purpose of lowering labor effort and saving experts time \citep{bejnordi2017diagnostic}. Moreover, the reliance on large, labeled datasets and the opacity of some deep learning models can also raise concerns about trustworthiness and explainability \citep{tajbakhsh2020embracing}, as it can't really be considered in sensitive situations unless they have a robust estimation of the uncertainty, so the doctors can determine when an additional examination is required;  this estimation is necessary due to the nondeterministic nature of the deep learning solutions, and a presence of noise or data shifting can drastically decrease the model performance \citep{brixtel2022whole, roy2024gru}.

As a result, there's been some interest in developing unsupervised segmentation algorithms.\citep{tajbakhsh2020embracing} developed self-learning techniques for nucleus segmentation, resulting in a dice score of 74.77\%. After that, \citep{javed2024unsupervised} proposed an unsupervised mutual transformer learning for WSI images.  While their results are encouraging, deep learning-based solutions need significant computational resources and frequently encounter trust concerns. To overcome these difficulties, various non-deep learning approaches have been offered. For instance, \citep{kochetov2024unseg} and \citep{zhang2024unsupervised} proposed non-deep learning nucleus segmentation with F1 scores of 66.70\% and 70-75\%, respectively. Although these methods are more computationally efficient and interpretable than deep learning algorithms, they fall short of the resilience and performance needed for widespread use in clinical practice. Moreover, one important disadvantage of these studies is their focus on nucleus segmentation, which is limited to high-resolution portions of the WSI image. Explainable unsupervised segmentation algorithms capable of analyzing both higher- and lower-resolution regions effectively (in which regions of interest (ROI) vary from tissues in lower resolutions to the nucleus in higher resolutions) have not been investigated to the best of our knowledge. This gap highlights the lack of a comprehensive unsupervised framework that is efficient, explainable, and effective across multiple resolutions in WSI.

We addressed the aforementioned limitation by designing an explainable, fully unsupervised, cross-level segmentation approach (TUMLS) for WSI images, which eliminates the need for annotation, reduces reliance on deep learning, and uses normalized distances from cluster centroids as an uncertainty-aware measure. It provides trustworthy, expressive insights, allowing experts to examine WSI images more quickly and confidently.

The rest of the paper is organized as follows: Section 2 showcases related work, including content-based image retrieval, histological segmentation, and classification. Section 3 describes the used datasets, covers the unique TUMLS framework, and provides experimental details. Finally, Section 4 presents the results, followed by a comprehensive discussion that outlines the strengths and limitations of the solution.

\raggedright
\section{Related works}
The use of machine learning (ML) for content-based image retrieval has been a topic of interest in recent research. \citep{ozturk2020stacked} introduced a stacked auto-encoder-based tagging approach with deep features for content-based medical image retrieval. \citep{guo2020privacy} focused on privacy-preserving image search (PPIS) using convolutional neural networks (CNN) for secure classification and searching over large-scale encrypted medical images. \citep{singh2021cbir} proposed a completely unsupervised four-convolution layer model for content-based image retrieval on celebrity data using deep CNNs. \citep{kanwal2020deep} presented a method that integrates symmetry, FAST scores, shape-based filtering, and spatial mapping with CNN for large-scale image retrieval, achieving highly accurate results in challenging datasets.
\citep{komura2022universal} explored universal encoding of pan-cancer histology through deep texture representations. \citep{camalan2020otomatch} developed a content-based image retrieval system for eardrum images using deep learning. \citep{zhuo2020low} proposed a low-dimensional discriminative representation method for high-resolution remote sensing image retrieval. \citep{deepak2020retrieval} focused on brain magnetic resonance imaging (MRI) retrieval with tumors using contrastive loss-based similarity on GoogLeNet encodings. \citep{kumar2021retracted} introduced a U-Net-based neural network for object-based image retrieval, showing improved performance in accuracy, precision, and recall. \citep{monowar2022autoret} introduced AutoRet, a self-supervised spatial recurrent network for content-based image retrieval. These studies collectively demonstrate the diverse applications and advancements in neural network methods for content-based image retrieval in various medical and non-medical domains.

Fully unsupervised segmentation leverages generally advanced machine learning techniques to analyze histology images without the need for extensive manual annotations. The proposed methodologies vary from weakly supervised, self-supervised, to fully unsupervised approaches. The authors of \citep{gadermayr2019generative} explored four techniques. The first method, SDS, employs a U-Net model for precise supervised segmentation. The second method alters the input image to a different stain before processing it through U-Net. This approach addresses the lack of annotations for each stain from the same sample. The third method, SDU, previously discussed in their earlier research \citep{gadermayr1805unsupervisedly}, involves creating artificial annotations. These annotations paired with input images create an optimal unpaired dataset for image-to-image translation using a cyclic GAN. The authors combined elements from the second and third methods, leading to a novel unsupervised approach for histology segmentation. One limitation of this study is that it uses artificially generated pictures based on visual preconceptions. The paper \citep{sebai2020maskmitosis} provided an RCNN-based deep learning framework for detecting mitosis in histopathology images, including fully supervised, weakly supervised, and unsupervised approaches. Besides, \citep{li2022net} proposed an unsupervised cancer segmentation framework for histopathology pictures utilizing the deep U-net architecture. The author used self-supervised learning (contrastive learning) for the data augmentation step, demonstrating that it is more efficient for model performance and complexity reduction than alternative augmentation strategies (at least in an unsupervised environment). Furthermore, \citep{van2024unsupervised} presented an unsupervised cell segmentation model that optimizes mutual information between data pairs, achieving accurate cell segmentation without expert annotations. Their model surpassed the SOTA of supervised and unsupervised models, with a Jaccard score of 0.71 on a proprietary dataset. Additionally, \citep{cisternino2024self} used self-supervised learning on a huge dataset of 1.7 million pictures to segment tissues and predict RNA expression. Their strategy surpassed existing methodologies, increasing the silhouette score by 43\%, correlating tissue morphology with gene expression. Furthermore, in the context of invariant information clustering, \citep{van2024unsupervised} suggested a model that optimizes mutual information for cell segmentation, achieving a Jaccard score of 0.71. This technique outperformed multiple supervised models, demonstrating its effectiveness in live-cell studies. Although previous studies successfully solved the constraint of data annotation, they continue to suffer from high computing resources and trustworthiness issues. Therefore, a quicker and more interpretable solution is proposed by \citep{zhang2024unsupervised}; its solution used several color-based transformations with k-means and fuzzy c-means (FCM) for nucleus segmentation.
On top of that, the comprehensive review \citep{wei2024overview} The comprehensive review \citep{wei2024overview} identifies several future gaps, including deep clustering interpretation, hyperparameter selection (DL and clustering parameters), multi-view deep clustering (as seen in \citep{chen2023deep}), and exploration of new domains. These findings inspire further research into superior algorithms for completely unsupervised histopathology segmentation.

In summary, content-based image retrieval studies influence high-level tissue segmentation; low-level nuclei segmentation is motivated by the achievements of \citep{kochetov2024unseg} and \citep{zhang2024unsupervised}; and pipeline explainability via high-level features and charts is inspired by \citep{REHAMNIA_TURLAPOV_2024}. Consequently, the main contributions are as follows:
\begin{itemize}
\item TUMLS provides concise, reliable cross-level results by highlighting relevant ROI from high and low resolution.
\item Quick segmentation algorithm that surpasses all unsupervised methods.
\item Only 2 levels encompass high and low levels, thereby diminishing redundancy and expediting the processing time.
\item An uncertainty-aware segmentation to improve the framework's reliability by offering robust predictive uncertainty estimation.
\item Eliminate the need for time-consuming manual annotations.
\item Mitigate the influence of human bias and the cost of decision.
\end{itemize}

\section{Methodology}
\subsection{Datasets}
In this study, we utilized the UPENN-GBM dataset, which comprises histological, MRI,... data from glioblastoma patients, as provided by \citep{bakas2022university}. Specifically, we focused on its histological section, which includes 71 whole slide images (WSI) in NDPI format. These slides are captured at 9 distinct magnification levels, though for computational efficiency, the OpenSlide Python library interprets them as 18 levels. Example patches extracted from these slides are shown in the figure below.

Given that the UPENN-GBM dataset is entirely unsupervised, an additional supervised dataset was required to evaluate the performance of the nucleus segmentation pipeline. For this purpose, we employed the MoNuSeg dataset, as introduced by \citep{kumar2017dataset}. This dataset contains well-annotated tissue images collected from multiple organs across various hospitals. The test set consists of 14 TIF images (2 histopathology samples from seven organs: breast, kidney, liver, prostate, bladder,colorectal, stomach) each of size 1000×1000×3 along with their corresponding annotations in XML format.

\subsection{Proposed Methodology}
In light of the relevant works above and in alignment with our objectives, the TUMLS methodology is given in Figure~\ref{methodology} below.

\begin{figure}[H]
\centering
\includegraphics[width=\textwidth]{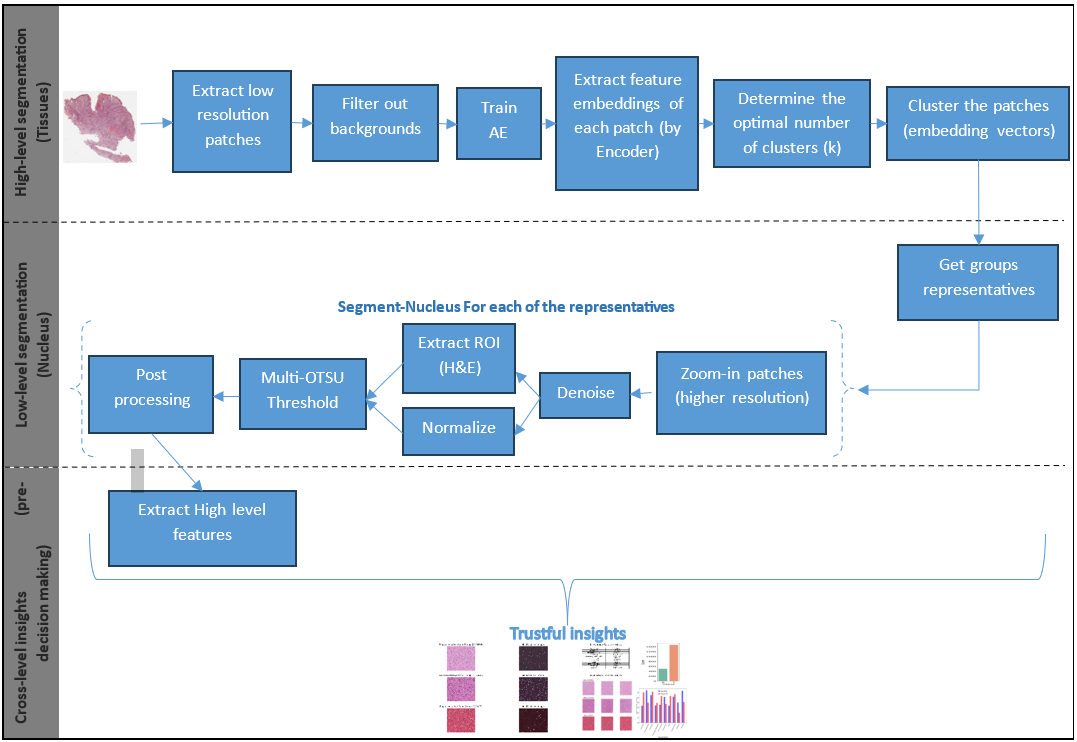} 
\caption{TUMLS Methodology}\label{methodology}
\end{figure}

\begin{itemize}
    \item \textbf{High-level segmentation (Tissues):}This phase recognizes existing tissues in the data and plots their representatives; it begins by breaking down a high level into smaller patches and filtering away the background patches based on statistical thresholds, after which the autoencoder (AE) is trained to reconstruct those patches. Following training, the encoder part of the AE is utilized to extract a 256 (2x2x64) latent space (feature embedding) for each patch. The patches are clustered with K-means based on these attributes (the hyperparameter k is determined by the optimal cut of the hierarchical clustering dendrogram). Finally, using the uncertainty-aware approach (based on distance to centroids), N best representative patches (closest to the centroid) for each tissue are provided.
    \item \textbf{Low-level segmentation (Nucleus):}This step performs nucleus segmentation on the relevant low level of the N representatives (from the previous phase, in which the fine details, such as the nucleus, are clearer). The approach begins with a denoising step that removes artifacts from the patches. The recovered hematoxylin channel is combined with the normalized patch (the normalization process is described in \citep{macenko2009method}). This combination is followed by multi-class OTSU thresholding, and final post-processing is handled using various morphological operations, resulting in 3 proposed masks: the first uses a closure followed by an opening, the second uses a custom nucleus demerging algorithm, and the third uses a median blur.
    \item \textbf{Cross-level insights (pre-decision making):} This phase extracts high-level morphological and texture features from the output of the previous pipelines (high and low-level ones) and compacts them with additional visualizations to yield the final insights that can potentially speed up the pathologist's decision-making process.
\end{itemize}

\subsection{Experimental Setup and Execution (Implementation details)}

Patch sizes and WSI levels are set empirically to provide spatial consistency across levels. To produce clear patches for the nucleus segmentation task, we start at the maximum resolution and use a patch size of 1024×1024 pixels. Any size larger than this resulted in slower processing times during the trial. We then decrease the resolution level by level until we reach level 12, at which point the small details vanish and only the tissue remains, which takes 6 steps. The patch size at level 12 is 16×16 pixels; it is obtained by substituting into Equation \ref{eq:size11}.

\begin{equation}
\text{Size}_{12} = \text{Size}_{18} \times \left( \frac{1}{2} \right)^{steps} \label{eq:size11}
\end{equation}

where:
$$\text{Size}_{18} = 1024 \times 1024$$
and
$$steps = 6$$

Substituting the values:

$$\text{Size}_{12} = 1024 \times 1024 \times \left( \frac{1}{2}\right)^{6}=16\times 16$$

The proposed framework is implemented in Python using the Pytorch framework on a personal computer with an NVIDIA GTX 1060 GPU.
To improve training efficiency in high-level segmentation, the encoder incorporates several layers from the pre-trained ResNet18 model (trained on the ImageNet dataset), whereas the decoder is a simple CNN-based decoder (with random initial weights). The AE network was trained utilizing the Adam optimizer with a batch size of 128, a learning rate of 0.001, and a weight decay of 1e-05, as well as batch normalization and early stopping (after 10 steps) as regularization strategies to improve accuracy and generalization.
Furthermore, we used an asynchronous patch prefetching mode with 12 workers (processes) to efficiently reduce the I/O bottleneck, considerably accelerating training (particularly with the weak GPU device and larger batch size) \citep{lin2018scannet}.

The illustrated low-level segmentation steps in the methodology above are supported by experimentation; for example, the experiment shows that multi-class Otsu's method is typically faster and much more resilient across different patches than K-means, which is slower and sometimes requires K hyperparameter tuning for better accuracy.

The UPENN-GBM dataset is divided into 3 subsets: training 70\% (124494 patches), validation 15\%(26678), and testing 1515\% (26678).
\subsection{Evaluation metrics}
\begin{itemize}
\item Mean Squared Error (MSE): a measure of the average squared difference between predicted and actual values. It estimates a model's prediction error for continuous outcomes. It is given by the formula~\ref{eq:mse} where a lower MSE suggests improved model performance \citep{james2013introduction}.
\begin{equation}
    \text{MSE} = \frac{1}{n} \sum_{i=1}^{n} (y_i - \hat{y}_i)^2
    \label{eq:mse}
\end{equation}
\item Dice Score (DSC): also known as F1 in the context of classification, it represents the harmonic mean of precision and recall, resulting in a balance between the two metrics. It is especially beneficial in cases where the class distribution is unequal. It is given by the formula~\ref{eq:dice} or formula~\ref{eq:f1} where a higher score suggests improved model performance \citep{sokolova2009systematic}.

\begin{equation}
\text{Dice} = \frac{2 |A \cap B|}{|A| + |B|} = \frac{\text{2 * TP}}{\text{2 * TP} + \text{FP} + \text{FN}}
    \label{eq:dice}
\end{equation}
or
\begin{equation}
F_{1} = 2\frac{\text{precision} * \text{recall}}{\text{precision} + \text{recall}} = \frac{\text{2 * TP}}{\text{2 * TP} + \text{FP} + \text{FN}}
\label{eq:f1}
\end{equation}
where:
\begin{itemize}
    \item $ |A| $ is the number of pixels in the predicted segmentation,
    \item $ |B| $ is the number of pixels in the ground truth segmentation,
    \item $ |A \cap B| $ is the number of pixels that are correctly predicted as belonging to the target class (nuclei).
    \item $ |TP| $ represents the true positive from the confusion matrix
    \item $ |FP| $ represents the false positive from the confusion matrix
    \item $ |FN| $ represents the false negative from the confusion matrix
\end{itemize}

\item Jaccard Similarity Coefficient (JSC): also known as intersection over union (IoU), is a statistic metric for determining the accuracy of an object identification model. It calculates the overlap between the expected bounding box and the ground truth box. Jaccard is calculated by dividing the area of overlap by the combined area of the predicted and ground truth boxes. It is given by the formula~\ref{eq:iou} where a greater IoU suggests improved model performance \citep{everingham2010pascal}.

\begin{equation}
\text{J(A,B)} = \frac{|A \cap B|}{|A \cup B|} 
\label{eq:iou}
\end{equation}

where:
\begin{itemize}
    \item $ |A| $ is the number of pixels in the predicted segmentation,
    \item $ |B| $ is the number of pixels in the ground truth segmentation,
    \item $ |A \cap B| $ is the number of pixels that are correctly predicted as belonging to the target class (nuclei),
    \item $ |A \cup B| $ is the total number of pixels in either the predicted segmentation or the ground truth segmentation (the union).
\end{itemize}

\end{itemize}

MSE is utilized to assess the reconstruction error of the proposed AE architecture in high-level segmentation, whereas dice and Jaccard are used to evaluate the generated nucleus masks in low-level segmentation.
\subsection{Extracted features}
Based on the Gray Level Co-occurrence Matrix (GLCM) and morphological properties of the segmented nucleus, the next 11 features are extracted: 
\begin{enumerate}[label=\arabic*-]
    \item Contrast: Measures the intensity contrast between a pixel and its neighbor over the whole image. High contrast indicates a greater difference between adjacent pixel values.
    \item Homogeneity: Assesses the closeness of the distribution of elements in the GLCM to the GLCM diagonal. High homogeneity means the pixel pairs are similar.

    \item Intensity: The average pixel intensity within the nucleus, reflecting the brightness.
    \item Staining Intensity: The degree of staining within the nucleus, often indicative of the concentration of a specific biomarker or dye.
    \item Number of Nuclei (num\_nuclei): The count of individual nuclei detected within the segmented area.
    \item Size: The area of the nuclei region measured in pixels.
    \item Circularity: A measure of how close the shape of the nucleus is to a perfect circle. Calculated by formula~\ref{eq:circularity} where a value of 1.0 indicates a perfect circle.
    
    \begin{equation}
        \text{Circularity} = \frac{4\pi \times \text{Area}}{\text{Perimeter}^2}
        \label{eq:circularity}
    \end{equation}

    \item Density: The number of nuclei per unit area, reflecting how tightly packed the nuclei are, is calculated by the next simple equation~\ref{eq:density}.
    
    \begin{equation}
        \text{Density} = \frac{\text{Number of Nuclei}}{\text{ROI Area}}
        \label{eq:density}
    \end{equation}
    
    \item Eccentricity: The ratio of the distance between the foci of the ellipse that best fits the nucleus to its major axis length. A value closer to 1 indicates an elongated shape, while 0 indicates a perfect circle; see the formula~\ref{eq:eccentricity}.

    \begin{equation}
        e = \sqrt{1 - \left(\frac{b}{a}\right)^2}
        \label{eq:eccentricity}
    \end{equation}
      Where:
        \begin{itemize}
            \item $ e $ = Eccentricity of the ellipse
            \item $ a $ = Length of the semi-major axis (the longest radius of the ellipse)
            \item $ b $ = Length of the semi-minor axis (the shortest radius of the ellipse)
        \end{itemize}  
    
    \item Spread: The spatial distribution of nuclei, describing how widely or narrowly they are dispersed within the area; it is calculated here as the standard deviation of the euclidian paiwise distances between nucleus centroids.

\end{enumerate}

\section{Experiments \& Discussion}

\subsection{High level (Tissues)}

\begin{figure}[H]
    \centering
\includegraphics[width=\textwidth]{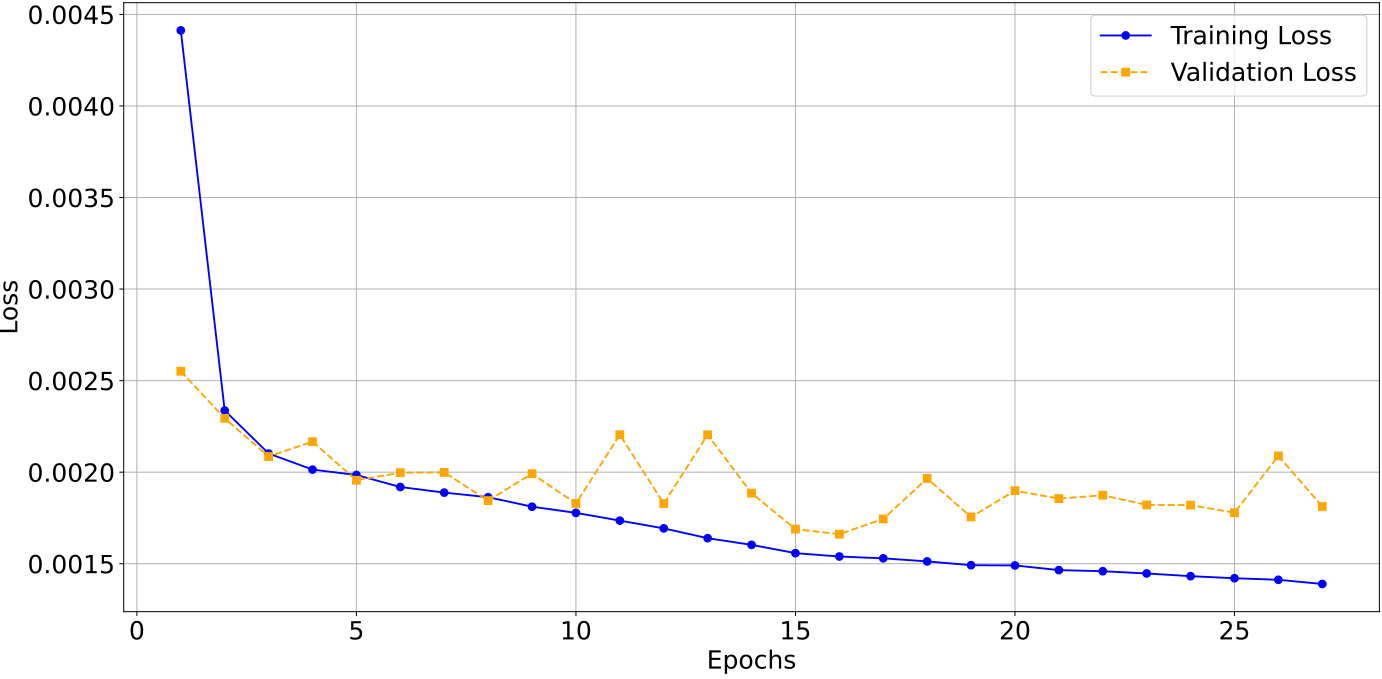}
    \caption{Training vs validation loss}\label{train_valid_losses}
\end{figure}

After 27 epochs, the training phase was terminated with the early stopping technique because the validation loss was no longer improving. Figure~\ref{train_valid_losses} shows that the 16th epoch yielded the best validation error of 0.00166, while the best training loss was the last one with 0.00138. The testing phase exhibits the same loss of validation.

\begin{figure}[htbp]
\centering
\includegraphics[width=\textwidth]{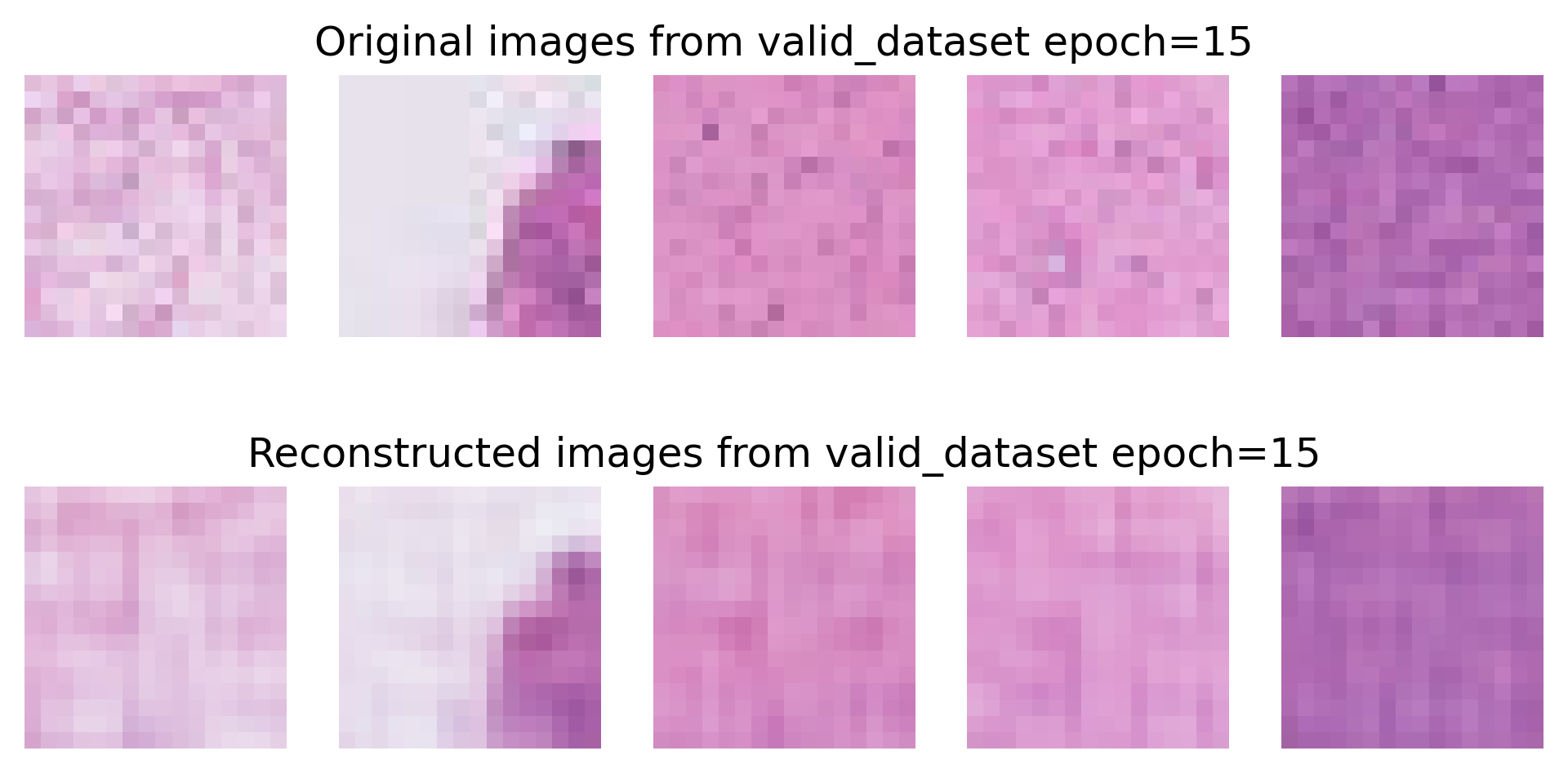} 
\caption{Comparison between original and reconstructed patches during training at the 15th epoch}\label{ae_reconstruction}
\end{figure}

During the training phase, whenever the validation error improves, a figure similar to Figure~\ref{ae_reconstruction}, which compares the original and reconstructed 5 random patches, is saved to visually assess the model's performance in capturing important tissue features.

\begin{figure}[H]
\centering
\includegraphics[width=\textwidth]{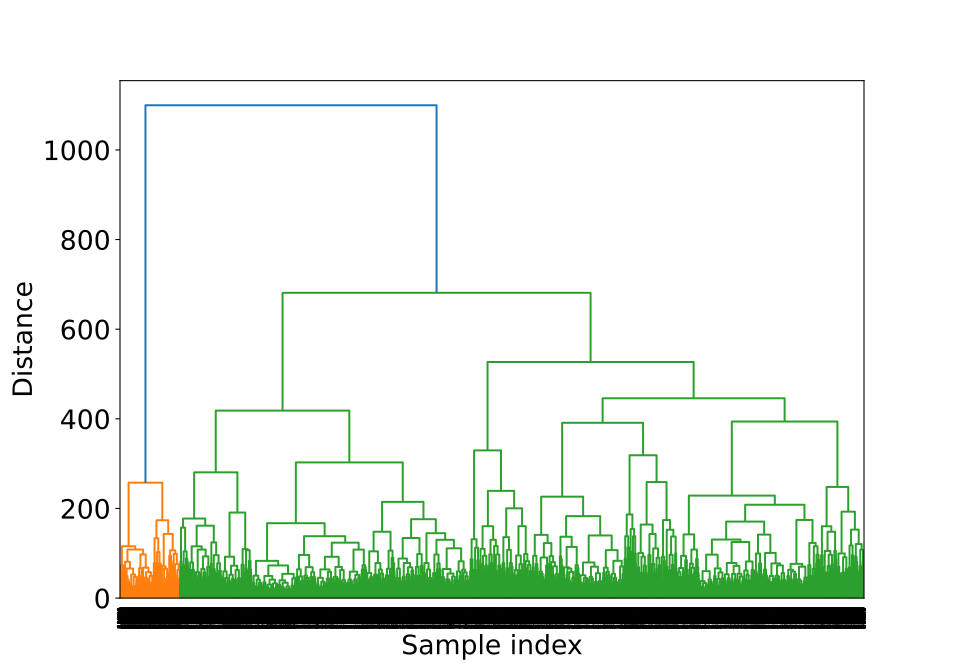}
\caption{Dendrogram of hierarchical clustering}\label{fig:dendogram}
\end{figure}

Following that, the deep clustering phase uses the flattened latent space formed by the encoder part of 256 (2x2x64) to identify tissues. As previously stated, a hierarchical clustering dendrogram is utilised to determine the appropriate number of groups using the best tree-cat technique, which involves selecting the cat with the largest distance in the dendrogram. Figure \ref{fig:dendogram} depicts an example with two tissues (longest distance).

\subsection{Low level (nucleus segmentation)}

\begin{figure}[htbp]
    \centering
\includegraphics[width=\textwidth]{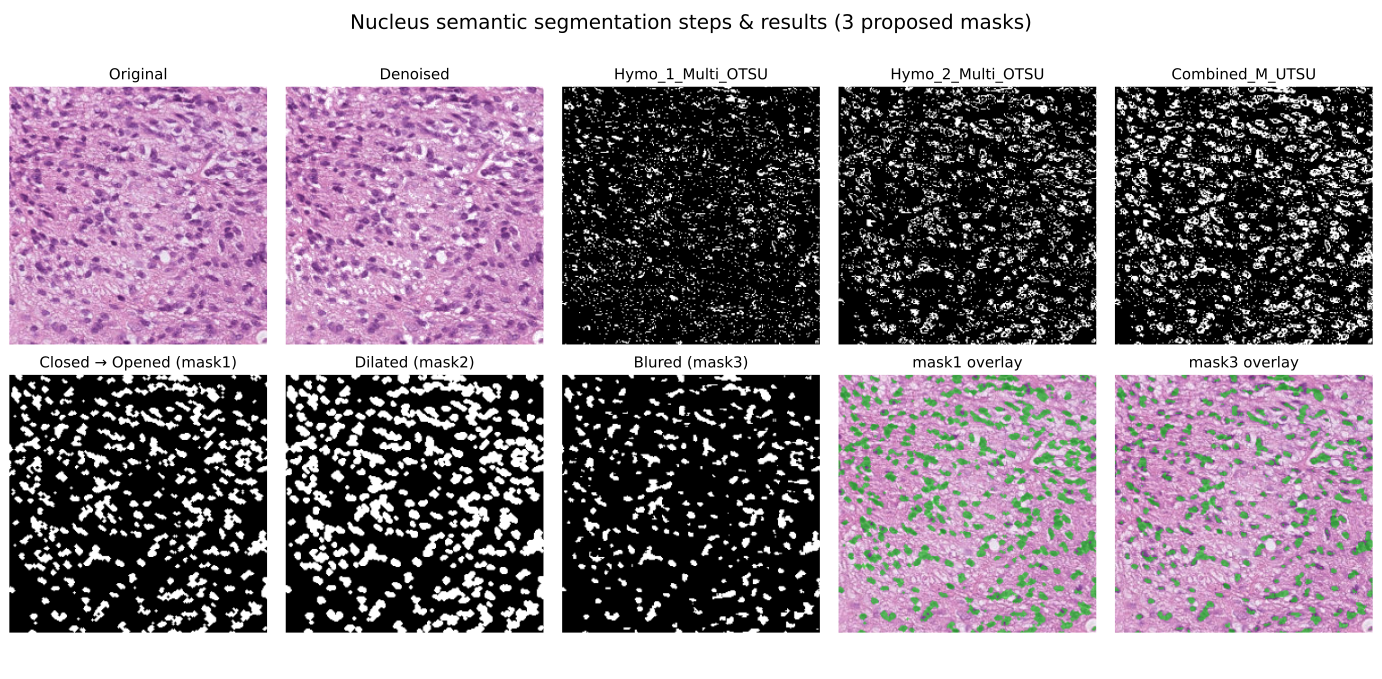}
    \caption{Nucleus segmentation steps visualized (without noise)}\label{fig:nucleus_segmentation}
\end{figure}

As explained above, low-level segmentation entailed many preprocessing steps and created 3 different masks. Figure~\ref{fig:nucleus_segmentation} shows the different preprocessing procedures in the ideal scenario (absence of noise).

\begin{figure}[H]
    \centering
\includegraphics[width=\textwidth]{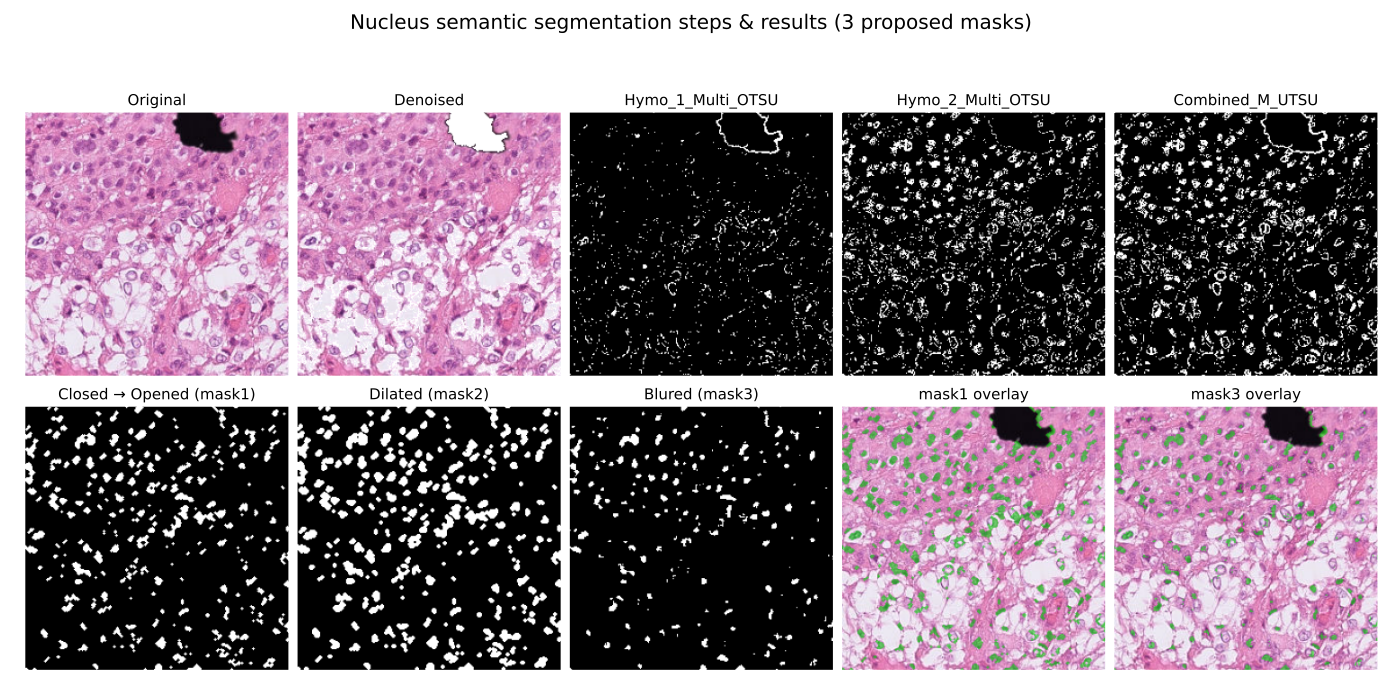}
    \caption{Nucleus segmentation steps visualized (with noise type 1)}\label{fig:nucleus_segmentation_n1}
\end{figure}

\begin{figure}[H]
    \centering
\includegraphics[width=\textwidth]{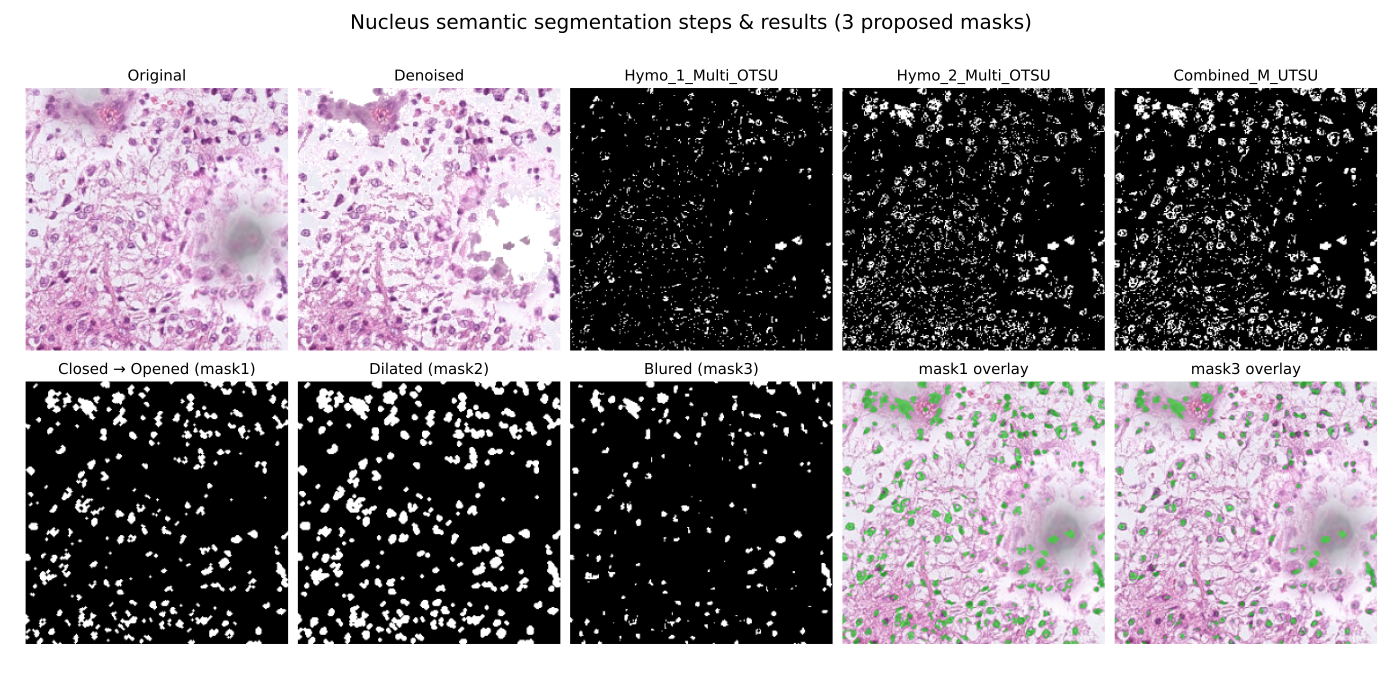}
    \caption{Nucleus segmentation steps visualized (with noise type 2 )}\label{fig:nucleus_segmentation_n2}
\end{figure}

Some WSI images from UPENN-GBM include significant noise, which is why a de-noising algorithm is performed on the input patch prior to any further processing. The analysis of patches discovered two forms of noise: a black likely region (Figure~\ref{fig:nucleus_segmentation_n1}) and a grey noise (Figure~\ref{fig:nucleus_segmentation_n2}).

Fortunately, the denoising operation removes both noises without sacrificing the important details in the image; more precisely, check at Figure~\ref{fig:nucleus_segmentation_n2} where the method successfully denoised the patch and found the two hidden nuclei. 
Some experiments are conducted to highlight the impact of its absence on further examination. The findings revealed that the algorithm tends to treat large noise regions as large nuclei, which has a significant impact on further analysis. Consequently, a non-tumorous area may be misclassified as tumorous due to the morphological characteristics associated with that noise.

\begin{figure}[H]
    \centering
\includegraphics[width=\textwidth]{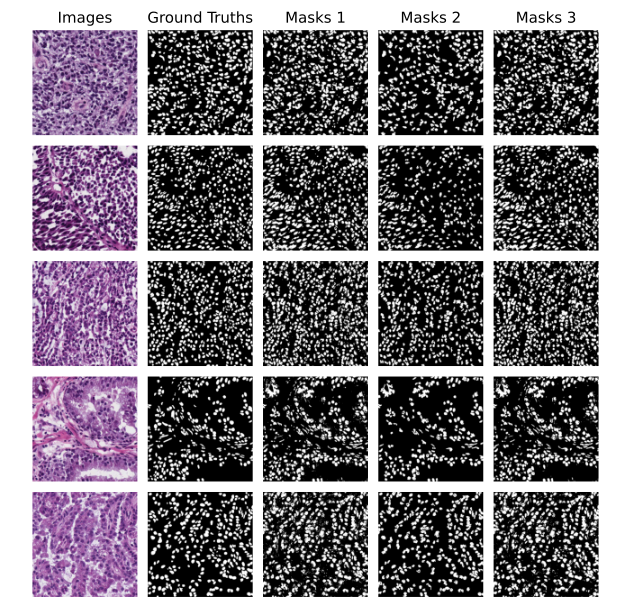}
    \caption{Comparison between the different generated masks for MoNuSeg dataset}\label{fig:monuseg_validation}
\end{figure}

While the results of nucleus segmentation on the UPENN-GBM dataset are promising, it is essential to compare this performance with that of state-of-the-art (SOTA) methods on a supervised dataset (MoNuSeg dataset). As a result, Figure~\ref{fig:monuseg_validation} shows the ground truth and several produced masks for different organs from the MoNuSeg dataset, and Table~\ref{tab:nucleus_segmentation_results_masks} displays the average performance for each of the masks, with mask 1 doing the best, slightly surpassing mask 3. Table~\ref{tab:nucleus_segmentation_per_organ} shows the performance of each of the 7 organs in the dataset. The brain was the best segmented organ, with an f1 score of 0.8159, while the breast organ had the lowest f1 score of 0.7418. Furthermore, whereas mask 1 is dominant in most organs, mask 3 slightly exceeds it in the lung and breast organs.

\begin{table}[H]
    \centering
    \begin{tabular}{|p{2cm}|p{2cm}|p{2cm}|p{2cm}|}
        \hline 
        \textbf{Organ}& \textbf{Mask 1} & \textbf{Mask2} & \textbf{Mask 3}\\ 
        \hline 
        Bladder  & 0.7943 & 0.7323 & 0.7622\\ 
        \hline
        Brain  & 0.8159 & 0.7696 & 0.8115\\ 
        \hline
        Breast  & 0.7418 & 0.6426 & 0.7426\\ 
        \hline
        Colon  & 0.7617 & 0.6408 & 0.7483\\ 
        \hline
        Kidney & 0.7842 & 0.6927 & 0.7595 \\ 
        \hline
        Lung & 0.7428 & 0.5957 & 0.7605 \\ 
        \hline
        Prostate & 0.7703 & 0.7074 & 0.7398 \\ 
        \hline
        \textbf{Overall} & \colorbox{yellow}{0.7746} & 0.6867 & 0.7614\\
        \hline
    \end{tabular}
    \caption{Detailed segmentation accuracy for each of the 7 organs of MoNuSeg dataset}
        \label{tab:nucleus_segmentation_per_organ}
\end{table}

\begin{table}[H]
    \centering
    \begin{tabular}{|p{2.4cm}|p{2.4cm}|p{2.9cm}|}
        \hline 
        \textbf{Mask type}& \textbf{F1 (Dice)} & \textbf{Jaccard (IoU)}\\ 
        \hline 
        1  & 0.7622 & 0.6168\\ 
        \hline
        2 & 0.6867 & 0.5270\\ 
        \hline
        3 & \colorbox{yellow}{0.7746}& \colorbox{yellow}{0.6335} \\ 
        \hline 
    \end{tabular}
    \caption{Average accuracy of the different generated masks}
        \label{tab:nucleus_segmentation_results_masks}
\end{table}

Moreover, Table~\ref{tab:nucleus_segmentation_comparison} demonstrates recent studies on nucleus segmentation using the MoNuSeg dataset, with the exception of one \citep{kochetov2024unseg}, which used a different evaluation dataset. The studies are sorted chronologically. Many of them do not include the Jaccard score, so the Dice score is used for comparison.


\begin{table}[H]
    \centering
    \resizebox{\textwidth}{!}{
    \begin{tabular}
        {|p{2.3cm}|p{7cm}|p{2.4cm}|p{1.1cm}|}
        \hline 
        \textbf{Study} & \textbf{Method} &\textbf{Learning approach} & \textbf{F1 (Dice)}\\ 
        \hline
        \citep{naylor2018segmentation} &  Mask R-CNN & Supervised & 74.70\\ 
        \hline
        
        \citep{naylor2018segmentation} &  U-Net & Supervised & 77.93\\ 
        \hline
                \citep{naylor2018segmentation} & FCN  & Supervised & 78.09\\ 
        \hline
                \citep{naylor2018segmentation} & DIST & Supervised & 78.63\\ 
        \hline

             \citep{tajbakhsh2020embracing} & Self-supervised approaches for nucleus segmentation & Unsupervised (Self-supervised) & 74.77\\ 
        \hline
        
        \citep{valanarasu2021medical} & LoGo:  Local-Global training strategy  & Supervised & 79.56\\ 
        \hline

        \citep{li2023lvit} & LViT-LW:  Language meets Vision Transformer in Medical Image Segmentation & Supervised & 80.66\\ 
        \hline

        
        \citep{roy2024gru} & GRU-Net: Supervised learning with Gaussian attention and dense skip connections & Supervised & 80.35\\ 
        \hline
        \citep{showrav2024hi} & Hi-gMISnet: Generalized segmentation using pGAN with dual-mode attention & Supervised & 82.5\\ 
        \hline 
       \citep{wang2024u} & U-HRMLP: Multi-scale feature fusion to refine segmentation boundaries & Supervised & 80.83\\ 
        \hline 
        \citep{wang2024narrowing} & UDTransNet:Learnable skip connections with DAT and DRA modules & Supervised & 79.47\\ 
        
        \hline 

        \citep{kochetov2024unseg} & UNSEG:Bayesian-like framework to segment cells and their nuclei without requiring training data & Unsupervised & 66.70 \\ 
        \hline
        \citep{zhang2024unsupervised} & YCbCr color space with FCM & Unsupervised & 70.99\\ 
        \hline
        \citep{zhang2024unsupervised} & YCbCr color space with k-means & Unsupervised & 75.53\\ 
        \hline
       The present study & TUMLS: Trustful Full Unsupervised Multi-Level Segmentation & Unsupervised & \colorbox{yellow}{77.46}\\
    \end{tabular}
    }
    \caption{Comparison of TUMLS with recent studies on nucleus segmentation}
        \label{tab:nucleus_segmentation_comparison}
\end{table}
\begin{itemize}
\item \textbf{TUMLS versus supervised approaches:} The f1 score of TUMLS is comparable to the 10 supervised approaches listed. It outperformed the Mask R-CNN strategy from \citep{naylor2018segmentation} by around 3\%, with a small drop in performance in the range of 0.5-5\% with the remainings.
The minor score decrease is due to its unsupervised nature, which does not require nucleus masks. This makes it an easy and efficient method to automatically segment nuclei.
Besides, according to  Table~\ref{tab:nucleus_segmentation_comparison}, 
deep learning (DL) solutions are highly effective in minimizing bias; however, they often struggle to maintain good variance across different datasets, resulting in solutions that are less generalizable. For instance, while the GRU-Net model in \citep{roy2024gru} outperformed SOTA solutions on the MoNuSeg dataset, achieving a dice score of 80.35\% and an IOU of 67.21\%, it experienced a significant drop in accuracy when trained on the “TNBC” dataset and tested on MoNuSeg, with dice declining to 65.98 and the IOU to 49.33\%. In contrast, the present study demonstrated greater resilience in maintaining accuracy. By conducting experiments on extracted patches from Whole Slide Images (WSI) and subsequently testing on the MoNuSeg dataset, our approach exhibited a competitive performance. This highlights the effectiveness of the proposed pipeline in achieving consistent results across varying datasets, addressing a critical limitation observed in existing DL methodologies.

\item \textbf{TUMLS versus unsupervised approaches: }The table shows that TUMLS outperformed all 5 unsupervised approaches. It outperforms the Bayesian-like foamwork UNSEG \citep{kochetov2024unseg} by approximately 11\%, the YCbCr color space with FCM \citep{zhang2024unsupervised} by around 7\%, the YCbCr color space with k-means \citep{zhang2024unsupervised} by about 2\%, and the self-supervised approaches \citep{tajbakhsh2020embracing} by about 3\%. Furthermore, TUMLS outperforms \citep{zhang2024unsupervised} solutions by eliminating the need to train any machine learning approach, reducing the burden of setting additional hyperparameters (K hyperparameters), and potentially improving performance. Experiments have demonstrated that the ideal k value varies by tissue type. Of course, the use of K-means in the high-level segmentation phase is obviously non-contradictory with this, since the k value is derived from the hierarchical clustering dendrogram. However, it should be noted that both of those clustering algorithms are based on the significant extracted feature vector rather than the full patch pixels, which contain a lot of noise and less significant information that could cause clustering errors.

\end{itemize}

\subsection{Cross-level insights}
The summarized insights can be generated either from a single WSI image or many, as follows:
\begin{itemize}
\item \textbf{N best representatives: }In this phase, the 5 most representative patches from each detected tissue are displayed in their respective low levels; the most representative are chosen based on the segmentation algorithm's uncertainty awareness, which is determined by the clustering algorithm's distance to the centroids (the matching low-level is determined using Python's "Open Slide" module). Figures ~\ref{fig:representatives_2}, ~\ref{fig:representatives_3}, and ~\ref{fig:representatives_4} show the top five representatives of the 2, 3, and 4 found tissues, respectively.

\begin{figure}[H]
    \centering
\includegraphics[width=\textwidth]{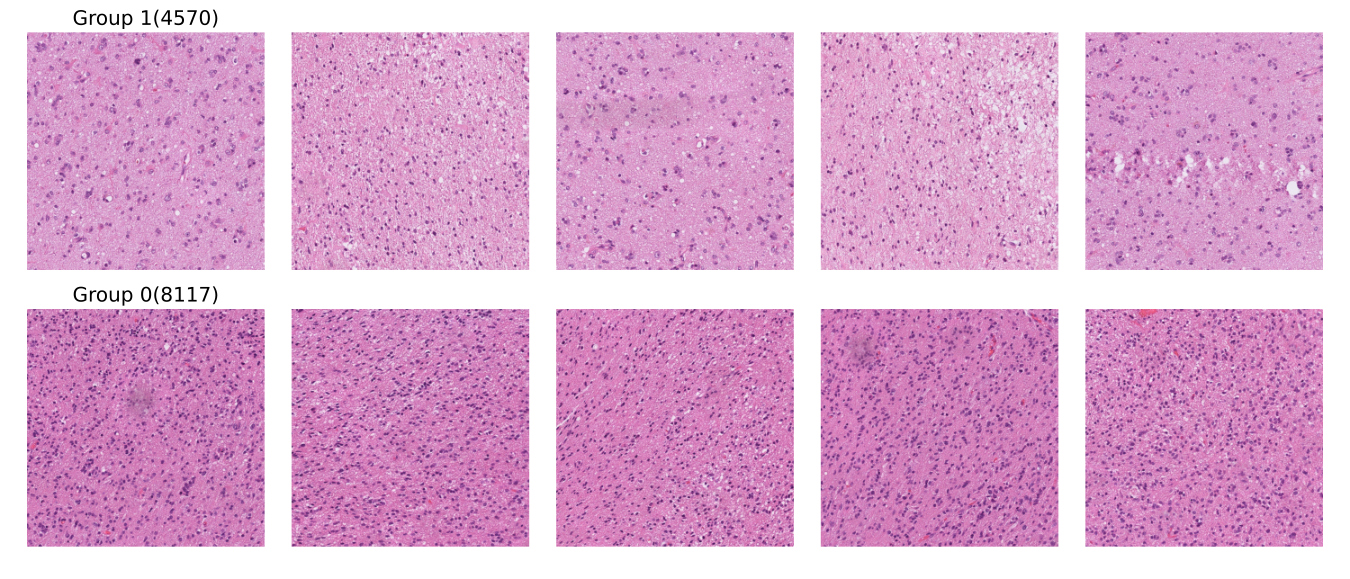}
    \caption{The best 5 representatives for each of the 2 detected tissues}\label{fig:representatives_2}
\end{figure}

\begin{figure}[H]
\centering
\includegraphics[width=\textwidth]{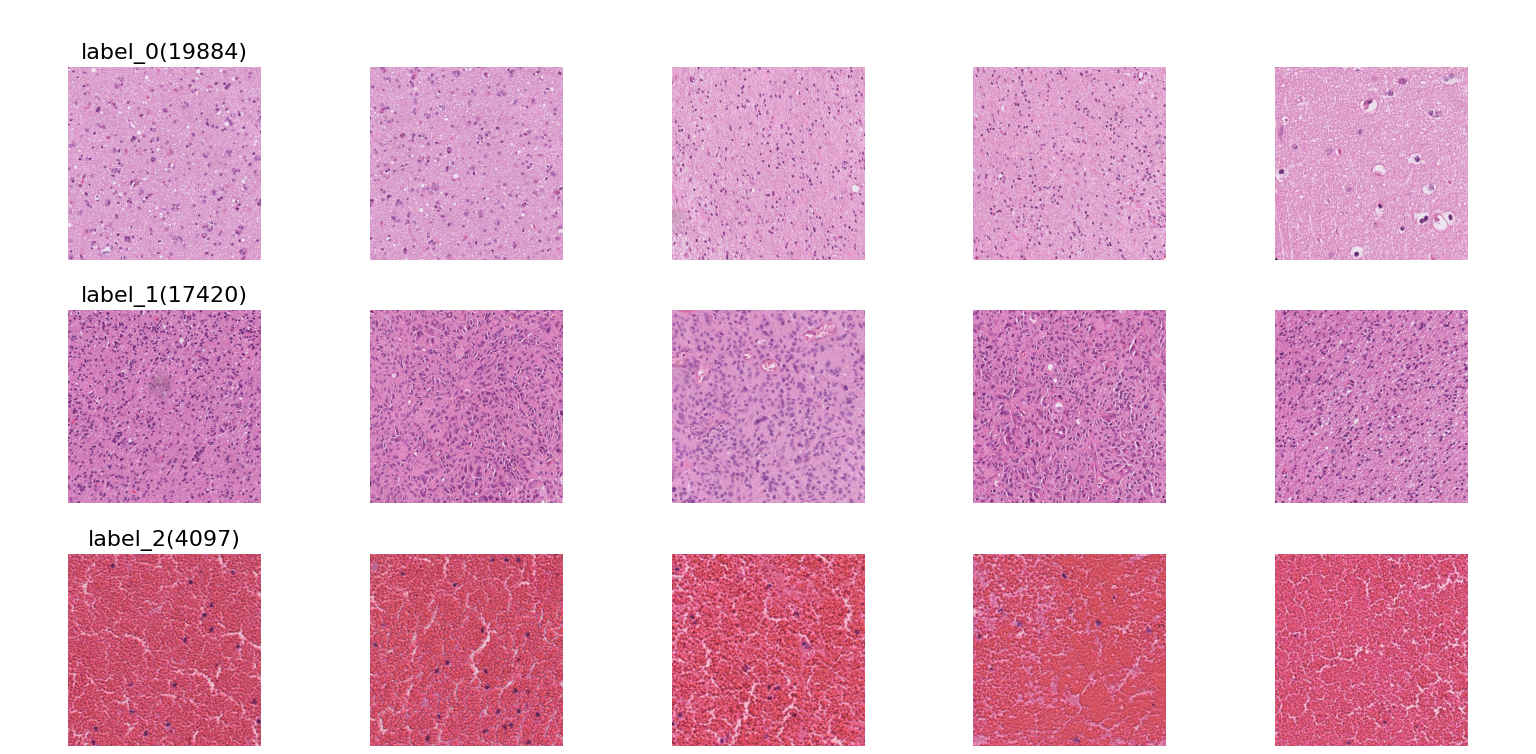} 
\caption{The best 5 representatives for each of the 3 detected tissues}\label{fig:representatives_3}
\end{figure}

\begin{figure}[H]
\centering
\includegraphics[width=0.8\textwidth]{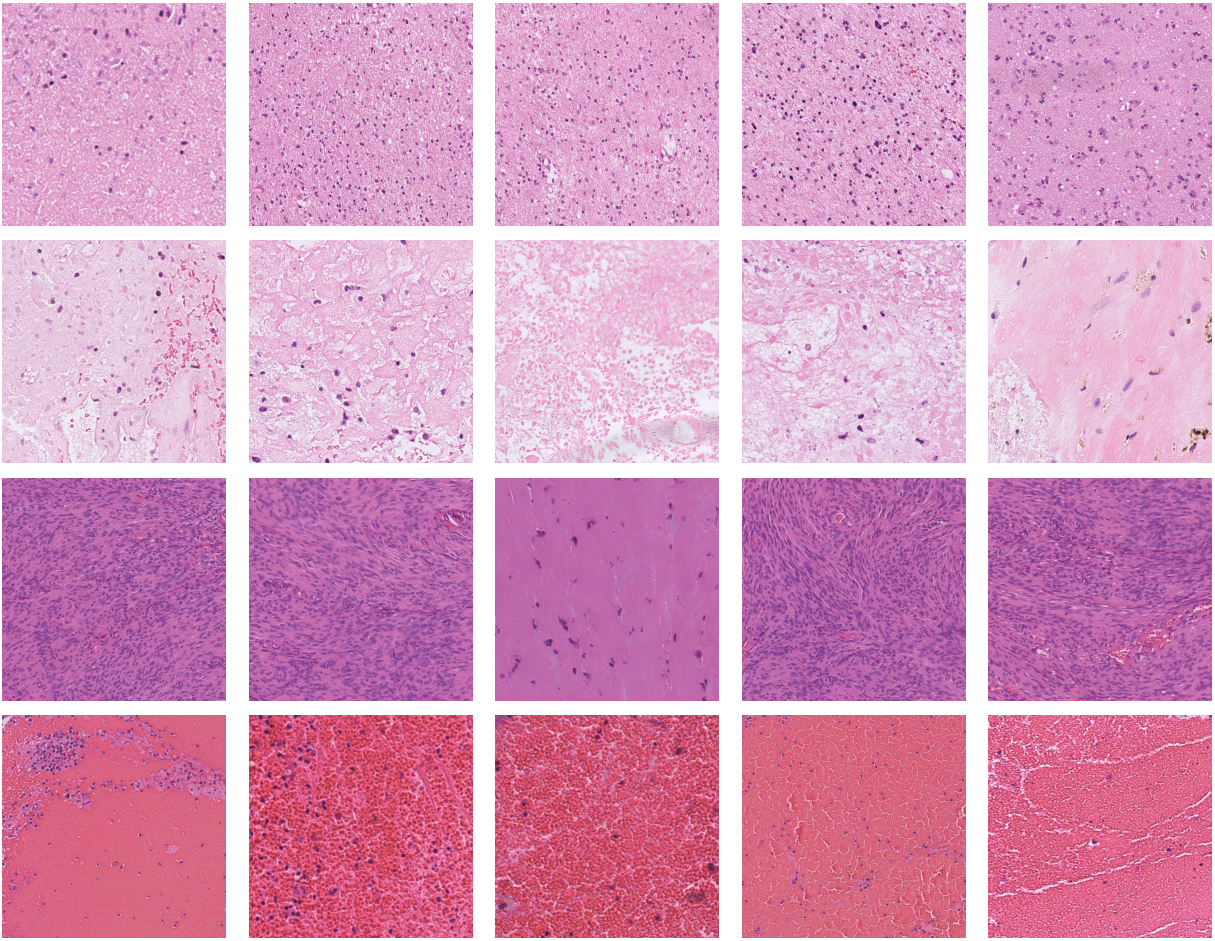} 
\caption{The best 5 representatives for each of the 4 detected tissues images}\label{fig:representatives_4}
\end{figure}

\item \textbf{Main insights: }In this phase, the best representative patch from each existing tissue (from the previous phase) is segmented into nuclei. The initial patch, the nucleus mask overlay on the original patch, and a listing of the main features specific to this tissue type are plotted. 
Figures~\ref{fig:insights_2} and ~\ref{fig:insights_3} display the top five cases discovered in the two and three tissues, respectively.

\begin{figure}[H]
    \centering
\includegraphics[width=\textwidth]{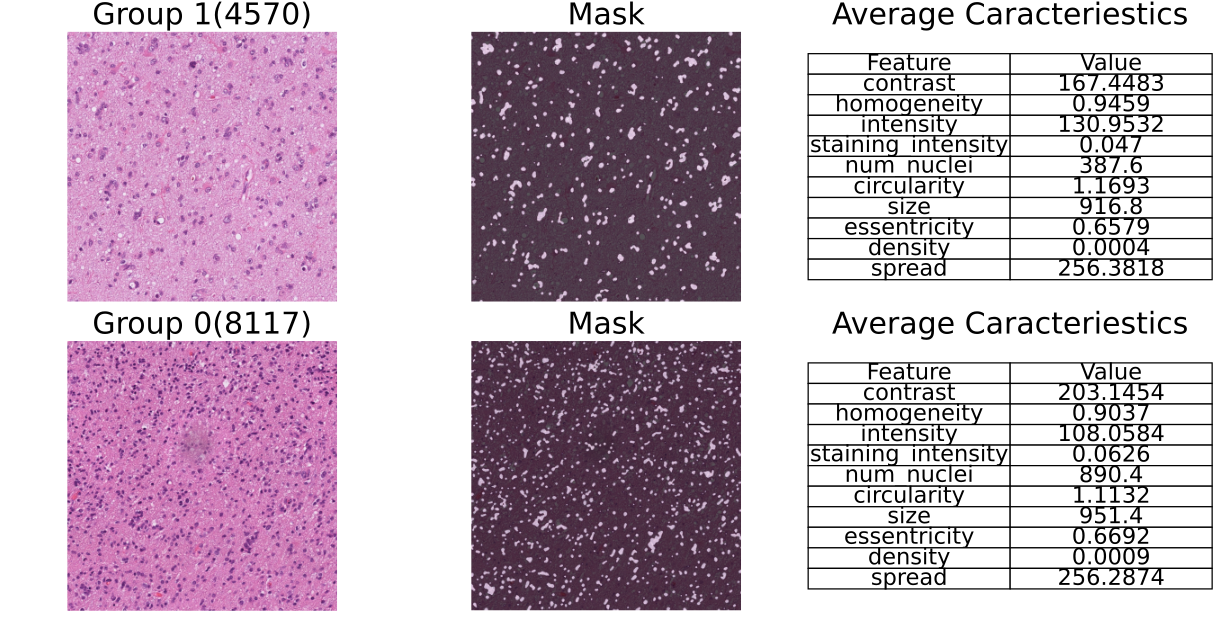}
    \caption{Insights for each of the 2 detected tissues image}\label{fig:insights_2}
\end{figure}

\begin{figure}[H]
\centering
\includegraphics[width=0.9\textwidth]{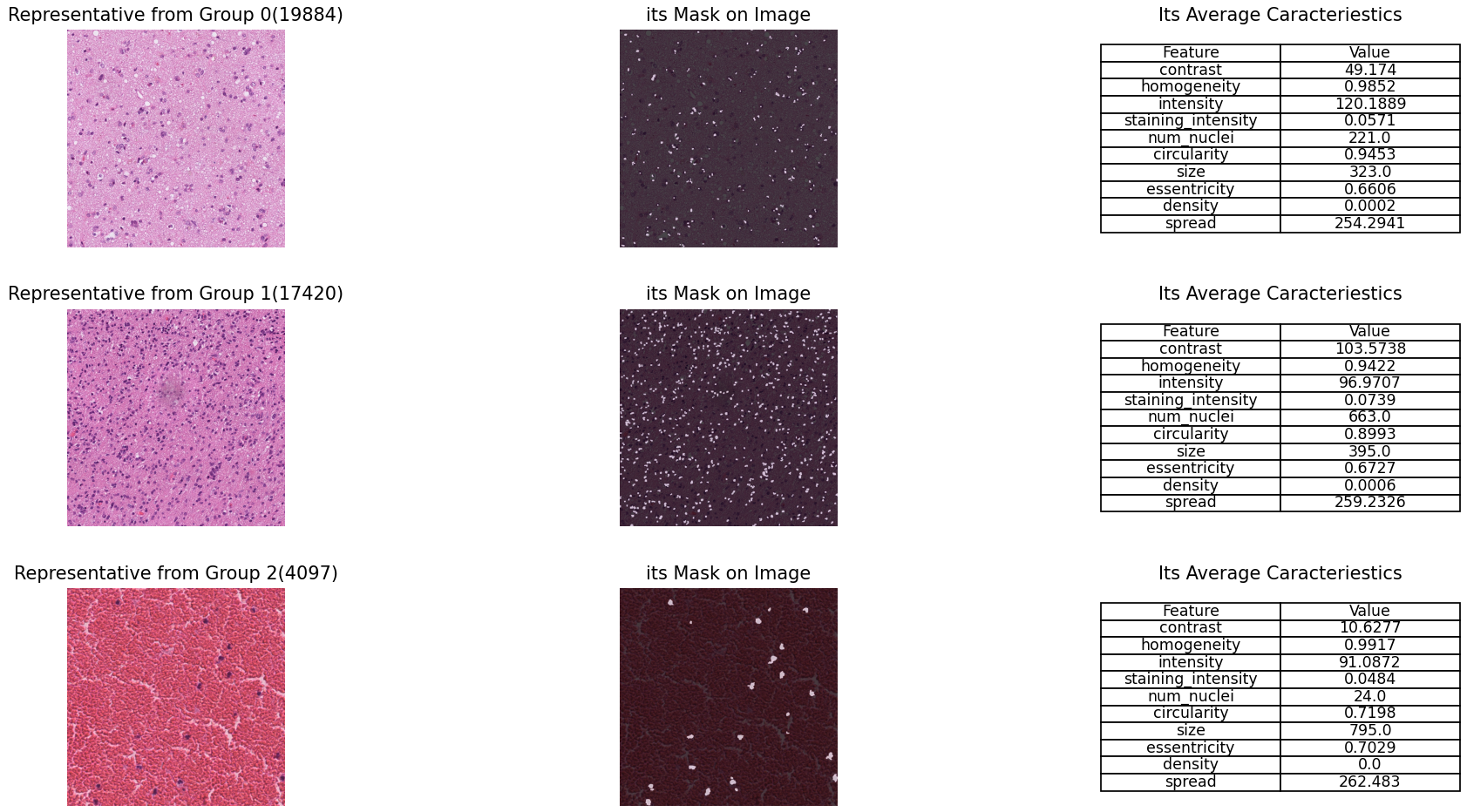} 
\caption{Insights for the 3 detetcted tissues}\label{fig:insights_3}
\end{figure}

\item \textbf{Comparison plots: }Figure~\ref{fig:comparison_histogram} shows a comparison of tissues based on their normalized features, while Figure~\ref{fig:tissue_distribution} displays the distribution of existing tissues across single or multiple WSI images (where 0 and 1 represent the ids of the first and second detected tissues, respectively).

\begin{figure}[H]
    \centering
\includegraphics[width=\textwidth]{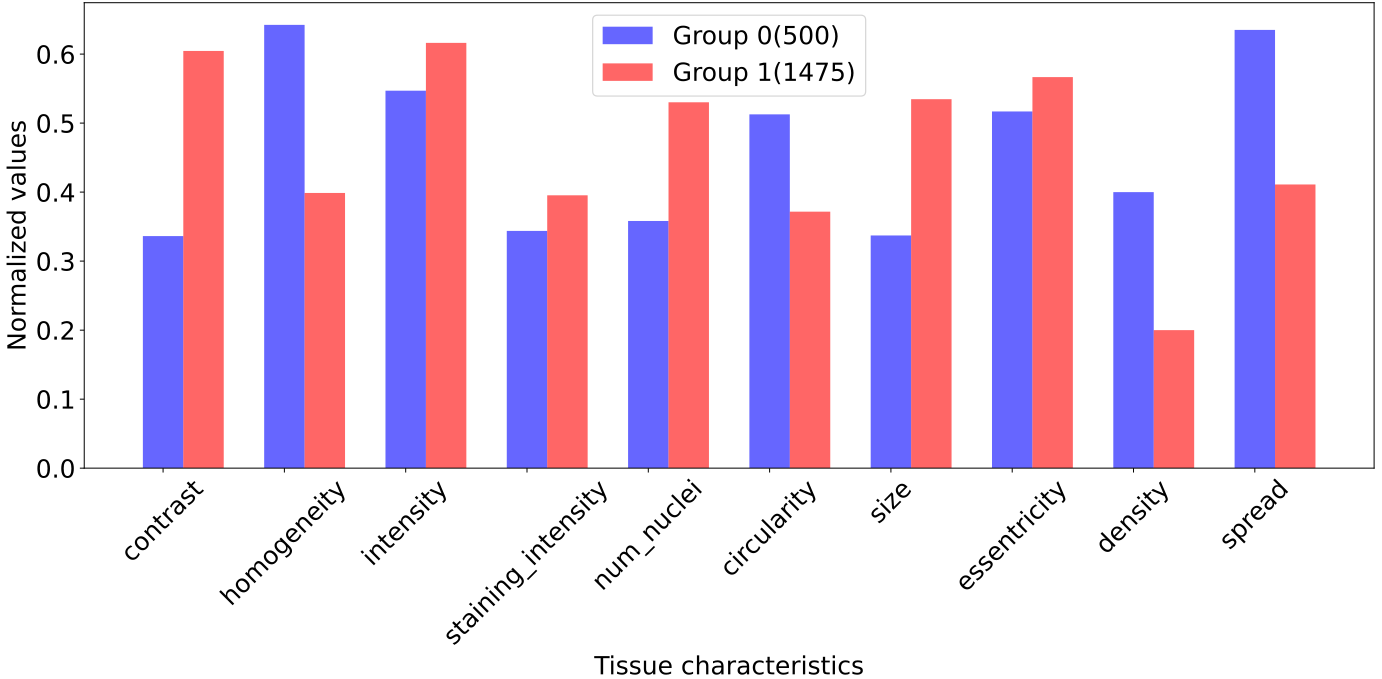}
    \caption{comparison histogram between the different tissues features}\label{fig:comparison_histogram}
\end{figure}

\begin{figure}[H]
    \centering
\includegraphics[width=0.4\textwidth]{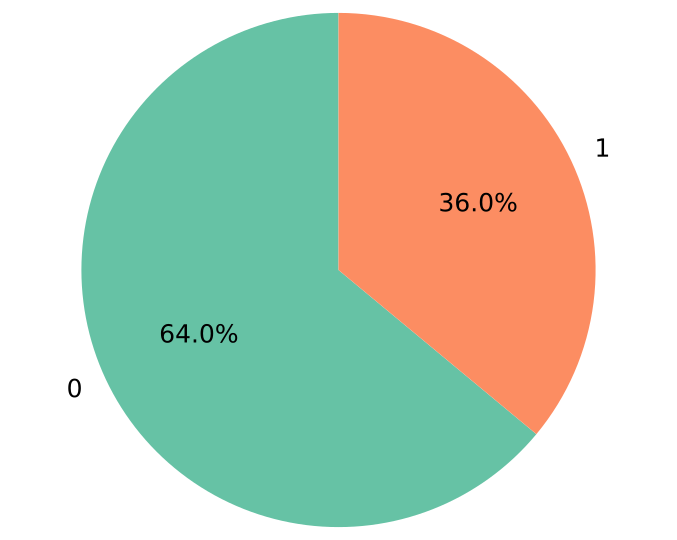}
    \caption{Tissue Distribution}\label{fig:tissue_distribution}
\end{figure}

\end{itemize}

Adopting the equivalent of best representatives at a lower level is intended to include cross-level ideas. As a result, the visual insights demonstrate TUMLS's efficiency and explainability on both high and low levels.

\subsection{Why this methodology?}
What makes the pre-decision insights of this solution reliable for clinicians are the next points:
\begin{enumerate}[label=\arabic*-]
    \item Annotation is completely unsupervised and does not require specialists.
    
    \item With a dice of 0.7746 and an IoU of 0.6335, this simple and quick nucleus segmentation outperforms all unsupervised systems and some supervised ones.
    
    \item The solution is trustworthy because it provides human control through pre-decision, safety through the uncertainty estimation approach, and explainability through high-level features and charts.
    
    \item Detecting existing tissues at a high level clearly directs the AE to focus on the coarser spatial characteristic while disregarding the finer one and accelerating the training process even more.
    
    \item Segmenting the nucleus without ML gives consistent accuracy across datasets (see its evaluation on the MoNuSeg dataset) and allows for operation at the lowest level 18 (closer to the nucleus) without significantly reducing pipeline performance.

    \item Using the AE-CNN-based technique, we can extract the most significant spatial information from the patches at a lower resolution of our choosing.

    \item Determining the k hyperparameter of K-means using the optimal cut of the hierarchical clustering dendrogram makes the solution more general and adaptive to the different cases.
    
    \item Sharing the same spatial vision in both high-level and low-level pipelines enables cross-level features and insights.
    
    \item Such a solution would significantly shorten the WSI investigation process without sacrificing regions of interest (ROIs) and important considerations.
    \item Knowing when to stop (pre-decision) and delegating vital decisions to specialists rather than anticipating them automatically (tumorous, healthy, etc.) is an important feature that most contemporary ML solutions lack. This feature prevents potentially disastrous decisions and boosts the methodology's credibility. 
    
    \item This AI solution does not replace or neglect the real specialists; rather, it allows them to make faster diagnoses.
    
    \item The absence of low-level features in the insights improves comprehension.
     \item The use of uncertainty measurements, such as normalized distance from the centroid, improves the reliability of the detected tissue type.
    \item Broad applicability: This methodology, unsurprisingly, allows for multi-level segmentation of various regions, which serves experts for a variety of purposes, including tissue segmentation, nucleus segmentation, ROI identification, investigation of regions surrounding the ROI, detailed morphological features of different tissues, and so on. 
    \item Generalization: This methodology applies to any type of tumor, not just brain tumors (as proved by the MoNuSeg dataset); it may also be used for other sorts of ROI in the medical profession.
    \item There is no redundancy; only one delegate from each of the existing groups.
\end{enumerate}

\subsection{Limitations}
\begin{enumerate}[label=\arabic*-]
    \item The segmentation pipeline is highly dependent on hematoxylin and eosin (H\&E) staining, which limits its effectiveness for unstained histological samples.
    
    \item The framework halts just before the final decision, offering both an advantage and a disadvantage; while this allows for greater flexibility, it also means the framework does not provide a complete end-to-end solution.
    
    \item The background filtering algorithm is not flawless, as it occasionally fails to exclude patches containing small edges of histological structures.
\end{enumerate}

\section{Conclusion}

This paper offers a reliable, entirely unsupervised, multi-level segmentation (TUMLS) method for WSI analysis. The TUMLS improves clinical workflow by eliminating annotations, lowering computational resource requirements while maintaining competitive performance, providing interpretable insights from only two representative levels, and incorporating uncertainty-aware measures into high-level segmentation.
Experiments show that the TUMLS is competitive despite its simplicity. We achieved a high-level reconstruction MSE error of 0.0016 and the best low-level performance among the unsupervised SOTA, with an F1 score  of 77.46\% and a Jaccard score of 63.35. Future work will focus on improving the accuracy of low-level nucleus segmentation and automating the selection of the most representative levels to optimize segmentation efficiency.

\section*{Code and Data Availability}
The data and code that support the findings of this study are available upon request.

\section*{Acknowledgement}
This work was supported by the Analytical Center for the Government of the Russian Federation (agreement identifier 000000D730324P540002, grant No 70-2023-001320 dated 27.12.2023).

\section{CRediT authorship contribution statement}
Vadim Turlapov: Supervision, Project administration, Conceptualization, Review, Editing, Funding acquisition.
Walid Rehamnia: Conceptualization, Methodology, Writing, Editing, Data curation and analysis, Visualization, Software and implementation.
Alexandra Getmanskaya: Review, Editing.


\bibliographystyle{elsarticle-harv} 
\bibliography{references}
\end{document}